\documentclass[epsf,useAMS,usenatbib,usegraphicx]{mn2e}
 
\usepackage{amssymb,amsmath}
\usepackage{epsfig}
\usepackage{rotating}

\title[Natal black-hole kicks in fallback supernovae]
{Natal kicks of stellar-mass black holes by asymmetric mass
ejection in fallback supernovae}
\author[Hans-Thomas~Janka]
  {Hans-Thomas~Janka\thanks{E-mail:
thj@mpa-garching.mpg.de}\\
  Max Planck Institute for Astrophysics, Karl-Schwarzschild-Str.~1, 
  85748 Garching, Germany
}

\date{\today}

\pagerange{\pageref{firstpage}--\pageref{lastpage}} \pubyear{0000}

\def\LaTeX{L\kern-.36em\raise.3ex\hbox{a}\kern-.15em
    T\kern-.1667em\lower.7ex\hbox{E}\kern-.125emX}

\usepackage[colorlinks=false]{hyperref}

\begin{document}

\label{firstpage}

\maketitle

\begin{abstract}
Integrating trajectories of low-mass X-ray binaries containing 
black holes within the Galactic potential,
Repetto, Davies \& Sigurdsson recently showed that the large
distances of some systems above the Galactic plane can only be
explained if black holes receive appreciable natal kicks. 
Surprisingly, they found that the distribution of black hole 
kick velocities (rather than that of the momenta) should be
similar to that of neutron stars.
Here I argue that this result can be understood if
neutron star and black hole kicks are a consequence of large-scale 
asymmetries created in the supernova ejecta by the explosion mechanism.
The corresponding anisotropic gravitational attraction of the
asymmetrically expelled matter does not only accelerate new-born
neutron stars by the ``gravitational tug-boat mechanism''. It can
also lead to delayed black-hole formation by asymmetric fallback
of the slowest parts of the initial ejecta onto the transiently existing 
neutron star, in course of which the momentum of the black hole
can grow with the fallback mass. Black hole kick velocities will
therefore not be reduced by the ratio of neutron star to black hole
mass as would be expected for kicks caused by anisotropic neutrino
emission of the nascent neutron star.
\end{abstract}


\section{Introduction}
Young neutron stars (NSs) are observed to possess space velocities 
in the range of $\sim$200--500\,km\,s$^{-1}$ on average (e.g.,
Lyne \& Lorimer 1994; Hansen \& Phinney 1997;
Arzoumanian, Chernoff \& Cordes 2002, Faucher-Gigu{\`e}re \& Kaspi 2006).
The fastest ones speed through interstellar space with velocities
in excess of 1000\,km\,s$^{-1}$. Such eigenmotions are much faster 
than typical velocities of Galactic stars and can also not
be explained by the breakup of binary systems as a consequence of the 
supernova (SN) explosion. This points to the compact remnants receiving 
natal kicks during their birth events. Natal NS kicks have also been
concluded from SN-remnant associations, from special characteristics
of the spin and orbital parameters of several 
compact binaries, and from evolutionary studies of the NS binary
population (e.g., Lai, Chernoff \& Cordes 2001; for a review, see
Lai 2001).

Observational evidence is also growing that stellar-mass
black holes (BHs) might receive 
recoil velocities at their formation, but the conclusions drawn from
the space velocities, orbital parameters, and locations of individual
binaries within the Galaxy are still ambiguous. While the space 
velocities of some systems do not seem to require natal BH kicks, 
e.g., Cygnus X-1 (Nelemans, Tauris \& van den Heuvel 1999) and
GRS~1915+105 (Dhawan et al.\ 2007), or are compatible with at most 
a small kick (Cygnus X-1; Wong, Willems \& Kalogera 2010), other
systems suggest that the kicks could be apprecible, e.g., in the  
cases of GRO~J1655-40 (Nova Sco; Brandt, Podsiadlowski \& Sigurdsson
1995; Willems et al.\ 2005) and XTE~J1118+480 (Remillard et al.\ 2000;
Mirabel et al.\ 2001; Gualandris et al.\ 2005; Fragos et al.\ 2009).

Recently, Repetto, Davies \& Sigurdsson (2012) considered the 
population of Galactic BH low-mass X-ray binaries (BH-LMXBs) 
as a whole instead of analysing the properties of individual systems.
They modeled the formation of BH-LMXBs in the Galaxy and applied
both mass-loss kicks of the binaries due to the SN explosion
of the primary star and additional 
natal kicks obtained by the newly formed BHs. Then they integrated
the trajectories of the binary systems within the Galactic potential.
Comparing the synthesized BH-LMXB population with the Galactic
distribution they concluded that birth kicks of the
BHs are {\em necessary} to explain the large distances above the 
Galactic plane achieved by some binaries. The hypothesis that BHs
only rarely receive a natal kick is ruled out with high significance.
Moreover, by comparing results for different theoretical kick
distributions, they found that BHs most likely have the same 
(or a very similar) distribution of kick velocities as NSs, which
means that the associated momenta of the BHs are higher by the ratio
of BH to NS mass.

This rather astonishing result can provide valuable
constraints of the physical mechanism that causes the recoil motions
of the compact SN remnants. Here it will be argued that two popular 
kick scenarios, namely neutrino-driven kicks caused by
anisotropic neutrino emission on the one hand and ``hydrodynamical''
kicks associated with asymmetric mass ejection in the SN blast
on the other hand
lead to opposite predictions. While anisotropic neutrino emission
must be expected to produce BH momenta of the same magnitude as
those of NSs, acceleration by explosion asymmetries can transfer
considerably larger momenta to BHs. It will be demonstrated that
in the latter case
the BH momentum can grow with the BH mass, which allows one to 
understand the similarity of the NS and BH velocity distributions
suggested by the analysis of Repetto et al.\ (2012).

When the explosion mechanism creates asphericities, the 
asymmetrically expanding ejecta mass exerts anisotropic, 
long-range gravitational forces on the compact remnant and thus
can lead to a long-lasting acceleration of the compact star. This was
shown with two-dimensional (2D) hydrodynamic simulations in the 
context of the neutrino-driven SN mechanism by
Scheck et al.\ (2004, 2006) and with three-dimensional (3D) SN
explosion models by Wongwathanarat, Janka \& M\"uller (2010, 2013), 
and it received independent confirmation through 2D explosion models 
of Nordhaus et al.\ (2010, 2012). Because the acceleration can
continue with considerable rates over a timescale of several seconds,
this ``gravitational tug-boat mechanism'' is much more efficient than
the impulsive hydrodynamical momentum transfer during the launch
phase of the explosion considered by Janka \& M\"uller (1994). 
The long-time gravitational acceleration can lead to NS 
velocities of at least more than 700\,km\,s$^{-1}$
(Wongwathanarat et al.\ 2013) and (in statistically probably less
frequent cases) has the potential to produce
even more than 1000\,km\,s$^{-1}$ (Scheck et al.\ 2006,
Wongwathanarat et al.\ 2013). 

The consequences of 
asymmetrical explosions for the kicks of BHs formed in fallback 
SNe will be discussed in Sect.~\ref{sec:bhgravkick}, 
after a brief description of neutrino induced NS and BH recoil
has been given in Sect.~\ref{sec:nukick} and a discussion
of NS acceleration by the gravitational tug-boat mechanism in 
Sect.~\ref{sec:nsgravkick}. A summary and conclusions will follow in
Sect.~\ref{sec:summary}.

\section{Neutrino induced NS and BH kicks}
\label{sec:nukick}

Anisotropic emission of neutrinos by the hot proto-NS (PNS) carries away 
momentum and can lead to a recoil acceleration of the NS. The energy
radiated in neutrinos equals the huge gravitational binding energy
of the NS; this amounts to 10--20\% of the 
rest-mass energy of the newly formed compact object, the exact value 
depending on still incompletely known properties of the supernuclear
equation of state (EoS), which determine the compactness of the remnant
(Lattimer \& Prakash 2001). Since
neutrinos escape with the speed of light ($c$), a small asymmetry of
the emission can account for appreciable NS kicks. For a radiated
neutrino energy $E_\nu$, corresponding to a radial momentum 
$p_\nu = E_\nu/c$, and for an emission asymmetry parameter $\alpha_\nu$,
the linear momentum transferred to the NS (opposite to the stronger
neutrino-emission direction) is
\begin{equation}
p_\mathrm{NS} = M_\mathrm{NS} v_\mathrm{NS} = \alpha_\nu \,
\frac{E_\nu}{c} \,.
\label{eq:nukick1}
\end{equation}
Introducing $f_\nu \sim 0.1$--0.2 as the ratio of the NS binding 
energy, $E_\nu = E_\mathrm{b}(M_\mathrm{NS})$,
to the NS rest-mass energy, $M_\mathrm{NS}c^2$, i.e.,
$E_\nu = E_\mathrm{b}(M_\mathrm{NS}) = f_\nu M_\mathrm{NS}c^2$, 
the NS velocity can be written as
\begin{equation}
v_\mathrm{NS} = \alpha_\nu\,f_\nu(M_\mathrm{NS})\,c
              = 300\,\left ( \frac{\alpha_\nu}{0.01}\right ) 
                     \left ( \frac{f_\nu}{0.1}\right ) \,
                     \frac{\mathrm{km}}{\mathrm{s}} \,.
\label{eq:nukick2}
\end{equation}
This means that an emission asymmetry $\alpha_\nu$ of only one percent,
$\alpha_\nu\sim 0.01$, could produce a NS kick around 300\,km\,s$^{-1}$. 

Because of the strong gravity of the NS, however, it is extremely 
difficult to explain a global anisotropy of even only one percent
that applies for the whole neutrino-cooling period of many seconds.
Convection inside the hot PNS (i.e., below the neutrinosphere)
cannot be an explanation, because the highly
time-dependent pattern of convective cells (whose angular diameter 
roughly equals the radial depth of the convective shell) stochastically
averages the associated asymmetries of the neutrinospheric emission.
The value of the corresponding effective neutrino-emission anisotropy
remains extremely small.

In order to define a nonstochastic, preferred direction of neutrino 
transport, a variety
of scenarios have been proposed, mostly involving ultra-strong 
magnetic fields ($>$$10^{15}$--$10^{16}$\,G) in the interior of the
nascent NS. If such fields could possess a strong dipolar component
(which may be questioned because of the violent, nonstationary,
high-multipolar convective mass motions in the hot PNS), or if they
could develop a broken mirror
symmetry by rapid differential rotation, this might lead to an
enhancement of the neutrino emission in one hemisphere,
e.g., either by the direct $B$-dependent
modifications of the neutrino opacities of nucleonic matter
(Bisnovatyi-Kogan 1996; Arras \& Lai 1999)
or by their impact on active-sterile neutrino-flavor oscillations
(e.g., Fuller et al.\ 2003; Kusenko 2009) or on the neutrino 
emission from a hypothetical quark phase in the core of the compact
remnant (Sagert \& Schaffner-Bielich 2008). 
All of these scenarios, however, invoke
more than a single ingredient of uncertain physics and must thus
be considered as highly speculative.

Nevertheless, neutrino-induced kicks by the mentioned scenarios
cannot be rigorously discarded and observational tests of the 
associated predictions are extremely desirable. In this context
the implications of the population of Galactic BH-LMXBs for BH 
kicks are very interesting.

BH formation in a stellar core-collapse event occurs after a
transient period of NS stability, during which the NS loses energy by
neutrino emission. The existence of such a phase
is crucial in the context of the neutrino-driven
explosion mechanism, where the neutrinos radiated by the NS deposit 
the energy that initiates and powers the expulsion of the outer stellar
layers in the SN. A BH eventually may form when the NS becomes 
gravitationally unstable due to a phase transition that softens the EoS
at supranuclear densities. Alternatively, instead of a reduction of the
maximum stable NS mass by such an EoS softening, the collapse to a BH
can be triggered by accretion of gas onto the NS until its mass 
exceeds the stability limit\footnote{The exact value of the 
maximum NS mass is currently not known because of our incomplete 
understanding of the supranuclear EoS. For a cold NS it is certainly above 
$(2.01\pm 0.04)\,M_\odot$ (Antoniadis et al.\ 2013) and probably lower
than $\sim$3\,$M_\odot$, which implies that the baryonic mass of the hot,
accreting object can be some ten percent higher before it becomes
gravitationally unstable.}. A SN explosion does not necessarily accompany
such a BH formation event\footnote{In the context of this paper the term
SN is used for stellar explosions driven by a successfully revived 
core-bounce shock. Low-velocity mass loss from the progenitor-star surface
in response to the reduced gravitating mass due to the neutrino emission
at the stellar center (Lovegrove \& Woosley 2013) has no relevance for the 
presented discussion.}. In the case of a successful explosion, however,
BH formation can happen by the later fallback of initially outward moving
matter that does not maintain a velocity larger than the escape
velocity to become unbound. Such a partial reimplosion of the star may
be triggered by reverse shocks which form in phases when the 
SN shock slows down and which decelerate the expanding stellar shells 
in the inner regions of the SN. 

When a BH forms via one of these paths the energy radiated in
neutrinos is bounded from above by the gravitational binding 
energy of the NS with the maximum possible mass:
$E_\nu \le E_\mathrm{b}(M_\mathrm{NS}^\mathrm{max}) = 
f_\nu(M_\mathrm{NS}^\mathrm{max}) M_\mathrm{NS}^\mathrm{max} c^2$. 
Using this in Eq.~(\ref{eq:nukick1}) and replacing there the 
NS mass and velocity by $M_\mathrm{BH}$ and $v_\mathrm{BH}$,
respectively, we find that the velocity of the BH is 
constrained by
\begin{equation}
v_\mathrm{BH} \le \alpha_\nu\,f_\nu(M_\mathrm{NS}^\mathrm{max})\,
\frac{M_\mathrm{NS}^\mathrm{max}}{M_\mathrm{BH}}\,\,c 
\le \frac{M_\mathrm{NS}^\mathrm{max}}{M_\mathrm{BH}}\,
v_\mathrm{NS}^\mathrm{max} \, .
\label{eq:nukick-bh}
\end{equation}
Since $f_\nu(M_\mathrm{NS})$ and $f_\nu(M_\mathrm{NS}^\mathrm{max})$ 
are of similar size and there is no reason why NSs that ultimately
collapse to a BH should radiate neutrinos with a systematically
larger asymmetry $\alpha_\nu$, Eq.~(\ref{eq:nukick-bh}) compared to
Eq.~(\ref{eq:nukick2}) implies that
BHs would receive kicks which are generally reduced by roughly a 
factor $M_\mathrm{NS}/M_\mathrm{BH}$ compared to those of 
NSs. This is in conflict with the results obtained by Repetto et 
al.\ (2012). We therefore conclude that observational evidence for
natal BH kicks larger than the maximum kicks of the reduced-velocity
distribution of NSs {\em disfavors the neutrino-induced
kick mechanism}, at least as an explanation of the BH recoil motions.
Moreover, the high velocities of BH-LMXBs also exclude that
the BHs have formed {\em without} associated SN explosions, because
in this case the natal BH kicks could only result from anisotropic 
neutrino emission and their velocities would have to be 
limited by Eq.~(\ref{eq:nukick-bh}).

In the sequence of arguments used above the assumption was made that 
the neutrino emission essentially stops once the BH has formed. 
This is a viable assumption as long as the fallback matter accreted
onto the newly formed BH has little angular momentum, because
matter that collapses essentially radially into a BH falls inward
too quickly to efficiently lose energy by radiating neutrinos.
With sufficiently large angular momentum, however, the fallback
matter could assemble into an accretion disk around the BH and
thus would continue to produce neutrinos at a significant rate. A 
corresponding hemispheric emission asymmetry would affect the
BH kick constraint of Eq.~(\ref{eq:nukick-bh}).

However, the angular momentum needed to keep matter on orbits
around a BH is huge. For Keplerian motion near the innermost
stable circular orbit, where the specific angular momentum $j$ has
an absolute minimum, one estimates that in the case of a Schwarzschild
BH $j_\mathrm{isco}(M_\mathrm{BH}) \gtrsim 5\times 
10^{16}\,(M_\mathrm{BH}/3\,M_\odot)$\,cm$^2$\,s$^{-1}$ is necessary.
Massive, BH forming stars, at least at solar metallicity, are 
unlikely to retain such amounts of angular momentum because of 
mass loss and the increased angular momentum loss mediated by 
magnetic torques (Heger, Woosley \& Spruit 2005; Langer 2012).
Similarly, also hydrodynamic instabilities during the pre-explosion 
phase like the spiral standing accretion shock instability
(Blondin \& Mezzacappa 2007; Fern{\'a}ndez 2010, Hanke et al.\ 2013)
are not likely to build up sufficient angular momentum in the 
region around the initial mass cut for stabilizing later fallback
material in a disk around the BH. The conclusion arrived at above
should therefore remain valid even if angular momentum plays a
role at some level.

\begin{figure*}
\vspace{10pt}
\begin{center}
\epsfxsize=0.6\columnwidth\epsfbox{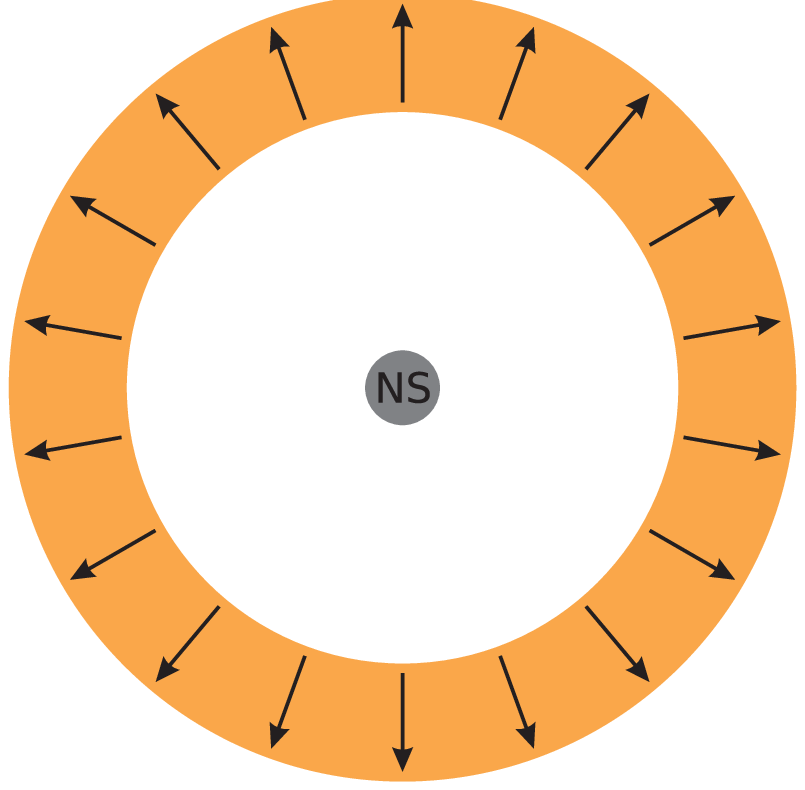}\hspace{25pt}
\epsfxsize=0.6\columnwidth\epsfbox{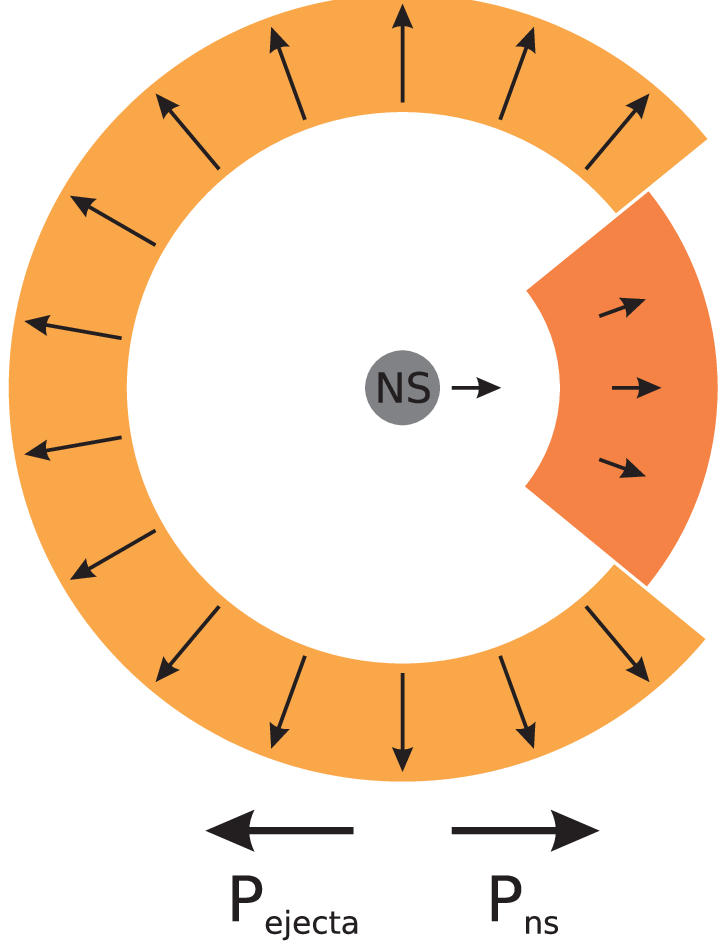}\hspace{25pt}
\epsfxsize=0.6\columnwidth\epsfbox{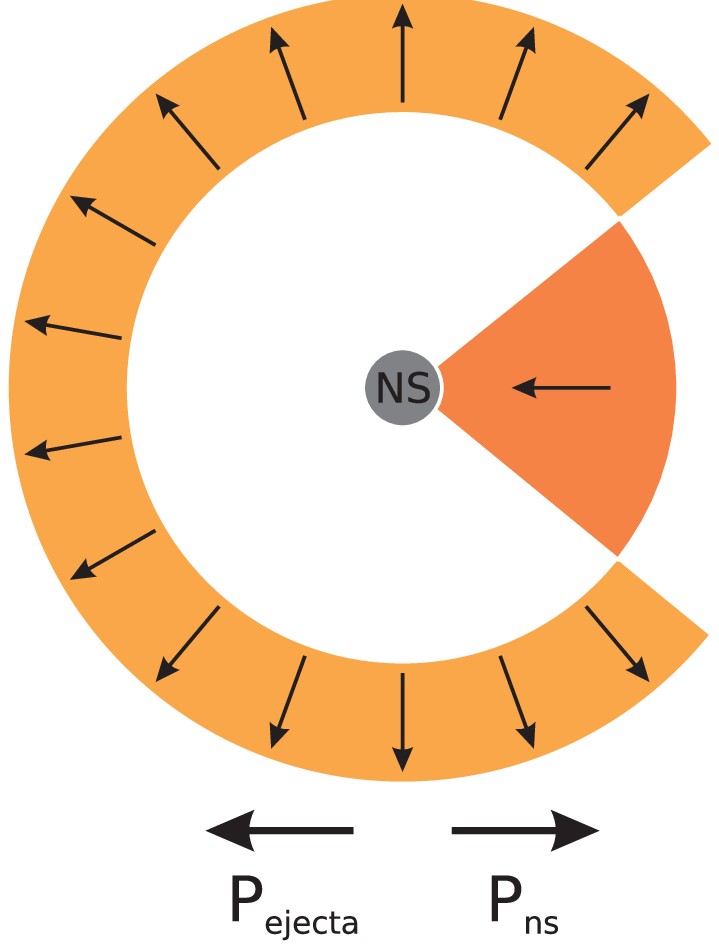}
\caption{Schematic visualization of SN mass ejection and compact
remnant kicks. In the left
image the ejecta are spherically symmetric and no recoil is imparted to
the central object. Asymmetric mass ejection must lead to compact remnant
motion with the opposite linear momentum (middle panel). 
The momentum can be transferred
by gravitational forces and by direct hydrodynamical forces in the case 
of accretion. The latter are crucial when the proto-neutron star is 
accreting fallback matter to collapse to a black hole (right panel). 
(Image taken from Scheck et al.\ (2006); reproduced with permission
\copyright ESO.)}
\label{fig:kickgeometry}
\end{center}
\end{figure*}

\section{Asymmetric mass ejection and compact remnant kicks}

The explosion mechanism of core-collapse SNe is nowadays
understood to be intrinsically multidimensional. This is suggested by 
observations, which require large-scale mixing and asymmetries
in the exploding stars (e.g., Leonard et al.\ 2006), 
and by theoretical considerations,
which fail to explain the onset of the SN blast with spherically
symmetric models except in the case of the lowest-mass 
(oxygen-neon-magnesium-core) progenitors (see, e.g.,
Janka 2012 and Burrows 2013 for reviews).
An asymmetric onset of the explosion will lead to anisotropic mass 
ejection. If associated with an anisotropic distribution of the
radial momentum in the outflow, total momentum conservation implies 
that the NS must obtain a kick with linear momentum opposite to that 
of the ejected mass in the rest frame of the progenitor star.

The exact mechanism that launches the SN shock wave
and creates the asymmetry of the energy and mass distribution
in the ejecta is not of crucial relevance; the underlying physics
of the NS acceleration is generally valid independent of
the specific processes that power the blast wave. 
The driving mechanism could be 
neutrino-energy deposition, magnetohydrodynamic effects, the acoustic
mechanism (for a recent review of these possibilities, see
Janka 2012 and references therein) or jittering jets 
(Papish \& Soker 2011),
and the mass-ejection anisotropy may not only be caused by hydrodynamic
instabilities that develop during the SN explosion but might also
result from pre-collapse asymmetries in the SN progenitor star
(Arnett \& Meakin 2011).
The transfer of momentum between ejecta and NS can happen directly
through hydrodynamic forces and indirectly through long-range
gravitational forces. These possibilities are schematically
illustrated in Fig.~\ref{fig:kickgeometry}.

\subsection{NS acceleration by the gravitational tug-boat mechanism}
\label{sec:nsgravkick}

In the context of neutrino-driven SN explosions, the 
acceleration of the newly formed NS by asymmetrical mass ejection
was demonstrated with  
2D hydrodynamic simulations by Scheck et al.\ (2004, 2006)
and further explored by Nordhaus et al.\ (2010, 2012). 
Simulations by Wongwathanarat et al.\ (2010, 2013)
have provided confirmation of the kick scenario also in three 
dimensions. A detailed discussion of the underlying physics
was provided by Wongwathanarat et al.\ (2013; see also Nordhaus
et al.\ 2010). The fundamental aspects of asymmetry creation and
NS repulsion are summarized here on the 
basis of the neutrino-heating mechanism but they are applicable
beyond this special case.

Seeded by potentially small initial perturbations in the
collapsing stellar core, large-scale asymmetries can develop by
hydrodynamic instabilities in the neutrino-heated accretion flow
between the stalled SN shock and the PNS at the center.
On the one hand, neutrino heating leads to convective overturn
and buoyant bubble motions (Bethe 1990), on the other hand global,
nonradial
shock deformation grows by the standing-accretion shock instability
(SASI; Blondin, Mezzacappa \& DeMarino 2003) with largest growth 
rates on the scale of the lowest-order spherical harmonics modes. 
The SASI is amplified in an advective-acoustic cycle (Foglizzo 2002,
Foglizzo et al.\ 2007, Scheck et al.\ 2008, Guilet \& Foglizzo 2012)
and leads to violent sloshing motions and
potentially spiral-mode rotation (Blondin \& Mezzacappa 2007,
Fern{\'a}ndez 2010, Hanke et al.\ 2013) of the postshock layer.

As a consequence, the mass and energy distributions around the
PNS can become highly asymmetrical with large variations of the 
neutrino-energy deposition rate in different directions. The explosion
sets in anisotropically with rising, high-entropy bubbles pushing 
the shock outwards while accretion to the PNS continues in narrower
downdrafts. Simultaneous accretion and shock expansion can persist
for periods of several 100\,ms, possibly in special cases 
for more than a second. 
Eventually, however, the outgoing shock accelerates the matter swept
up in the overlying shells to velocities exceeding the local escape
speed, and the inward pointing accretion downdrafts lose their 
supply of fresh gas. Accretion to the PNS therefore decays and 
the transition to the neutrino-driven wind phase takes place, 
during which the PNS continues to lose mass at a low and declining rate
($\lesssim$10$^{-3}$...10$^{-2}$\,$M_\odot$\,s$^{-1}$) in an essentially 
spherically symmetric supersonic outflow. This outflow is driven by the 
energy that neutrinos escaping from the hot PNS deposit in the layers 
adjacent to the neutrinosphere (Duncan, Shapiro \& Wasserman 1986;
Qian \& Woosley 1996).

Accretion downflows and rising bubbles can directly transfer momentum
to the NS by hydrodynamic effects (Janka \& M\"uller 1994). 
Because of the strongly time dependent
and nonstationary flows before and during the initiation of the 
explosion, however, the NS is bounced back and forth in varying 
directions and can attain only small recoil velocities (at most around
100\,km\,s$^{-1}$) in multi-dimensional simulations of this early 
postbounce phase (e.g., Fryer 2004; Fryer \& Young 2007). 
Once the explosion sets in and the asymmetry pattern of the 
ejecta becomes frozen in, however, significant linear momentum 
in a certain direction can build up in the compact remnant
(as visualized in the right panel of Fig.~\ref{fig:kickgeometry}).
This NS recoil points away from the direction of strongest mass ejection.
In contrast to such a hydrodynamic acceleration associated with the 
accretion downdrafts and anisotropic outflows, the spherically
symmetric neutrino-driven wind does not contribute to the NS recoil
on any significant level.
Instead, the compact remnant experiences a persistent traction
that is exerted by the anisotropic gravitational attraction of the
asymmetrically distributed ejecta. The corresponding NS acceleration
is directed towards the densest, slowest ejecta clumps
(Fig.~\ref{fig:kickgeometry}, middle panel). The NS velocity grows at
the expense of ejecta momentum, but the associated loss
of momentum in the outward moving ejecta can be compensated by their
continuous reacceleration as internal explosion energy is converted to 
kinetic energy by hydrodynamic forces in the accelerating SN blast. 
The NS and the towing high-density ejecta clumps can also be considered
as a mass entity that moves jointly in a shared gravitational trough
and interacts hydrodynamically (and gravitationally) with the 
surrounding SN material, similar to a sailing ship that is blown
by the wind and is pulling a dinghy on a tow line.

The long-distance gravitational coupling between compact remnant
and dense, slowly expanding ejecta is the most efficient contribution
to the long-lasting acceleration of the newly formed NS. In 3D 
simulations this gravitational tug-boat mechanism was shown
to be able to accelerate the NS to velocities in excess of
700\,km\,s$^{-1}$ over timescales of many seconds (Wongwathanarat et 
al.\ 2010, 2013), and analytic estimates (Wongwathanarat et al.\ 2013)
as well as 2D simulations (Scheck et al.\ 2006) suggest that in 
extreme cases more than 1000\,km\,s$^{-1}$ are well possible.

\subsection{BH acceleration in fallback supernovae}
\label{sec:bhgravkick}

Asymmetric mass ejection and the associated acceleration of the 
compact remnant by hydrodynamical and gravitational interaction
has interesting implications in the case of BH formation in 
fallback SNe, i.e.\ in SN explosions where the 
energy injected to the blast wave is not sufficient to unbind
the whole stellar mantle and envelope. In this case a larger
fraction of the progenitor will fall back to the newly formed
NS after its initial expansion has been slowed down
by reverse shocks that originate 
when the outgoing SN shock wave decelerates in regions where the 
density gradient is more shallow than $\rho \propto r^{-3}$
(e.g., Kifonidis et al.\ 2003). If sufficiently massive,
the fallback will trigger the delayed collapse of the NS 
to a BH. The BH in LMXBs must have gained its mass by such a
massive fallback in the course of the SN explosion of the
primary star.

As described in Sect.~\ref{sec:nsgravkick}, the PNS in an asymmetric
explosion is accelerated mainly due to the gravitational attraction
of the slowest-moving parts of the ejecta. The gravitational
interaction between compact remnant and the innermost ejecta 
also determines the fallback in the SN. Therefore the 
fallback is likely to occur preferentially in the directions 
where the explosion is weaker, whereas the matter with the
highest expansion velocities has the best chance to escape from 
the gravitational influence of the central object. This anisotropic
fallback will enhance the momentum asymmetry of the SN mass 
ejection, thus {\em increasing} the kick velocity of the compact remnant.
According to the arguments given in Sect.~\ref{sec:nsgravkick},
the corresponding additional acceleration of the accretor is 
mediated by hydrodynamical and gravitational forces associated
with the anisotropic infall of the fallback 
matter (Fig.~\ref{fig:kickgeometry},
right panel). Since the asymmetry of the fallback should
correlate with the momentum asymmetry of the initially (i.e., after
the saturation of the NS kick and prior to the fallback) expanding
SN matter, the momentum of the BH forming remnant must be expected
to increase with the fallback mass. From this scenario one therefore
expects a BH momentum distribution that grows with the BH mass,
or, in other words, a velocity distribution of BHs which may be
similar to that of NSs and which is {\em not} reduced by the ratio of
NS to BH mass.

In order to make the argument more transparent and formal, let us 
consider a simple toy model, in which the initial explosion
ejecta have a hemispheric asymmetry with mass $m_+$ and average
velocity $v_+$ in one hemisphere and $m_-$ and $v_-$ in the other
such that $0 < v_- \le v_+$ and $m_- \le m_+$ (see
Fig.~\ref{fig:kickgeometry}, middle panel). We can then write for
the momenta carried (in opposite directions) by the ejecta and by 
the recoiled NS:
\begin{equation}
p_\mathrm{ej} = p_\mathrm{NS} = m_+v_+ - m_-v_-
              = \alpha\, \bar{v}_\mathrm{ej}\, M_\mathrm{ej} \,.
\label{eq:ejectamom}
\end{equation}
Here $M_\mathrm{ej} = m_+ + m_-$ is the total ejecta mass,
$\bar{v}_\mathrm{ej} = (m_+v_+ + m_-v_-)/M_\mathrm{ej}$ the average
ejecta velocity, and $\alpha = (m_+v_+ - m_-v_-)/(m_+v_+ + m_-v_-)$
the momentum asymmetry of the ejecta.

Let us now assume that a mass $M_\mathrm{fb}$ falls back to the
compact remnant and is added to the initial PNS mass, $M_\mathrm{NS}$,
to yield a BH mass of
\begin{equation}
M_\mathrm{BH} = M_\mathrm{NS} + M_\mathrm{fb}\, ,
\label{eq:bhmass}
\end{equation}
which grows linearly with the accreted mass. 
If the fallback mainly affects the most slowly expanding
ejecta (as reasoned above), the initial ejecta component
$m_-$ is reduced to $m_-' = m_- - M_\mathrm{fb}$ (primed quantities
denote the state after the fallback, unprimed ones correspond to the
conditions prior to the fallback).
The momenta of the expelled SN matter and of the BH
after the fallback are:
\begin{equation}
p_\mathrm{ej}' = p_\mathrm{BH} = m_+v_+ - (m_- - M_\mathrm{fb})\,v_-
               = p_\mathrm{NS} + M_\mathrm{fb}\, v_- \, .
\label{eq:fbkick}
\end{equation}
This means that not only the mass of the BH but also the BH momentum 
increases with the fallback mass and that the BH momentum exceeds that
of the predecessor NS. Although the ejecta mass of the SN 
is reduced when gas is gravitationally captured again by the accreting 
compact object, i.e., $M_\mathrm{ej}' = M_\mathrm{ej} - M_\mathrm{fb}$,
the asymmetry of the mass ejection as well as the 
average velocity of the ultimately escaping gas will grow
(Fig.~\ref{fig:kickgeometry}, right panel). Linearizing the 
corresponding quantities for $M_\mathrm{fb}/M_\mathrm{ej} \ll 1$
(which implies even an underestimation of the extreme 
conditions when the fallback affects a major fraction of the
stellar mass) one obtains for the momentum asymmetry parameter:
\begin{equation}
\alpha' \sim \alpha + (1+\alpha)\,\frac{v_-}{\bar{v}_\mathrm{ej}}\,
\frac{M_\mathrm{fb}}{M_\mathrm{ej}} \, ,
\label{eq:fbalpha}
\end{equation}
and for the average ejecta velocity:
\begin{equation}
\bar{v}_\mathrm{ej}' \sim \bar{v}_\mathrm{ej} + (\bar{v}_\mathrm{ej}-
v_-)\,\frac{M_\mathrm{fb}}{M_\mathrm{ej}} \, .
\label{eq:fbejvel}
\end{equation}
Since $\bar{v}_\mathrm{ej}- v_- > 0$ the correction terms dependent
on $M_\mathrm{fb}$ in both Eqs.~(\ref{eq:fbalpha}) and 
(\ref{eq:fbejvel}) are positive, which means that the primed quantities
are larger than the unprimed ones.

Combining Eq.~(\ref{eq:ejectamom}) ($p_\mathrm{ej} = p_\mathrm{NS}$)
with Eqs.~(\ref{eq:bhmass}) and (\ref{eq:fbkick}), one obtains for
the BH kick velocity
\begin{equation}
v_\mathrm{BH} = \frac{M_\mathrm{NS}}{M_\mathrm{BH}}\,v_\mathrm{NS} +
\frac{M_\mathrm{fb}}{M_\mathrm{BH}}\, v_- \approx
\frac{M_\mathrm{NS}}{M_\mathrm{BH}}\,v_\mathrm{NS} + v_- \,,
\label{eq:bhfbvel}
\end{equation}
where the transformation for the inequality makes use of 
$M_\mathrm{BH} \sim M_\mathrm{fb} \gg M_\mathrm{NS}$.
Equation~(\ref{eq:bhfbvel}) compared to Eq.~(\ref{eq:nukick-bh})
constitutes the main difference between BH kicks as a consequence of
explosion asymmetries in fallback SNe and BH kicks associated with
NS kicks caused by anisotropic neutrino emission. While the latter
are limited by the maximum velocity that NSs can receive, reduced
by a factor of NS to BH mass, explosion asymmetries do not imply
such a constraint for the BH velocities. The fallback-SN
scenario for accelerating stellar BHs is therefore compatible with
the large natal BH kicks concluded by Repetto et al.\ (2012)
from their population synthesis analysis of the observed distribution
of Galactic BH-LMXBs. 

It should be noted, however, that in the discussed scenario the 
distribution of natal BH recoil velocities is not predicted to be
the same as the natal NS velocity distribution. If $M_\mathrm{BH}
\gg M_\mathrm{NS}$, the second term depending on $v_-$ in
Eq.~(\ref{eq:bhfbvel}) yields the dominant contribution to the
BH velocity $v_\mathrm{BH}$. The 
distribution of BH kick velocities therefore needs to be 
determined by a large set of 3D supernova simulations for a wide
variety of progenitor stars, for which the explosion asymmetry and
the long-time evolution of the anisotropic fallback are
followed over periods of possibly days. While such
simulations are currently not available, one can still estimate
the magnitude of the corresponding BH kick velocities.

In order to become gravitationally unbound, SN debris has to expand
with velocities in excess of the escape velocity, otherwise it will
be gravitationally captured by the central compact remnant. Fallback
will therefore affect matter whose velocity after deceleration by
a reverse shock becomes less than the escape velocity, i.e., if 
$v_-' < v_\mathrm{esc}$. Since the passage through a reverse shock
decelerates the expansion velocity by a factor that is roughly equal
to the inverse of the density jump in the reverse shock, 
$v_-/v_-' \approx \rho'/\rho \equiv \beta$ with $4\le \beta\le 7$ 
depending on the equation of state of the SN gas, the condition for 
initial ejecta being swallowed by later fallback can be written as:
\begin{eqnarray}
v_- \!\!\!&<&\!\!\! \beta\,v_\mathrm{esc}(r) \approx \beta\,\sqrt{2GM(r)/r}
                                            \nonumber \\
    \!\!\!&\sim&\!\!\! 1130\,\left (\frac{\beta}{4}\right )
                    \left (\frac{M}{3\,M_\odot} \right )^{\! 1/2}
                    \left (\frac{r}{10^{12}\,\mathrm{cm}}\right )^{\! -1/2}
                    \,\,\frac{\mathrm{km}}{\mathrm{s}} \,.
\label{eq:escapevel}
\end{eqnarray}
Here $\beta = 4$ for a nonrelativistic ideal gas was used, 
a value of 3\,$M_\odot$ corresponding to a ``seed BH mass'' was
chosen for the normalisation of the gravitating mass, and a radius
of $10^{12}$\,cm constitutes an approximate lower bound of the radial 
scale where the strong reverse shock from the
He/H composition interface begins to affect the ejecta expansion (see,
e.g., Kifonidis et al.\ 2003). The result of Eq.~(\ref{eq:escapevel}) 
shows that extreme conditions (namely, highly asymmetric fallback
as assumed in the sequence of steps leading to Eq.~\ref{eq:bhfbvel})
are able to produce BH kick velocities that can compete with the greatest
measured space velocities of NSs. BH kicks of several 100\,km\,s$^{-1}$
appear easily compatible with anisotropic fallback in SNe.

\section{Summary and conclusions}
\label{sec:summary}

I have argued that explosion asymmetries created by the SN mechanism
are likely to explain not
only the natal kicks of NSs but also those of BHs. Clumpy and inhomogeneous
ejecta exert long-duration, anisotropic gravitational forces on the NS and
can thus accelerate the compact object over timescales of seconds to 
velocities of many hundred kilometers per second
(Scheck et al.\ 2004, 2006; Nordhaus et al.\ 2010, 2012;
Wongwathanarat et al.\ 2010, 2013). Similarly, the slowest-moving parts
of the ejecta, especially when reverse shocks decelerate the expanding
SN flow, can be gravitationally recaptured by the NS. The anisotropic
fallback of such SN matter may not only trigger the collapse of the
NS to a BH but also allows the momentum of the compact remnant to increase
roughly proportionally to the accreted (and thus BH) mass 
(Eqs.~\ref{eq:bhmass} and \ref{eq:fbkick}), 
and the BH kick velocities can become similarly large as those of 
NSs (Eqs.~\ref{eq:bhfbvel} and \ref{eq:escapevel}).
This fallback-SN scenario for BH formation and recoil is therefore 
compatible with the results of Repetto et al.\ (2012), according to
which large BH natal kicks are required by the wide spread of the
observed distribution of BH-LMXBs around the Galactic plane.

In contrast,
momentum-conserving BH kicks, for which the BH velocities would be
reduced by a factor $M_\mathrm{NS}/M_\mathrm{BH}$ compared to NS
recoil velocities, are ruled out by the study of Repetto et al.\ (2012)
with high statistical significance. This strongly disfavors the 
possibility that BH kicks are simply inherited from their predecessor NSs,
because such a connection would constrain the BH momentum by the 
maximum possible NS momentum (Eq.~\ref{eq:nukick-bh}).
In particular, it disfavors scenarios, in which intrinsic NS properties
like anisotropic neutrino emission caused by strong magnetic dipolar
fields are responsible for the NS acceleration without asymmetric mass
ejection playing any important role for the kick. In a special
variant of this scenario (assuming a powerful neutrino-induced NS kick
early after core bounce), Fryer \& Kusenko (2006) found that the 
neutrino-heating mechanism initiates 
a stronger mass ejection of the SN explosion
{\em in the direction} of the recoil motion of the NS. Such a hypothetical
situation, however, appears to be in conflict with the results of 
Repetto et al.\ (2012). Because fallback must preferentially occur on
the weaker side of the explosion, the increasing momentum of the
remaining ejecta can only be balanced if the BH that forms by the fallback
accretion is accelerated {\em opposite to} the movement of the NS.
This effect must be expected to grow with the initial NS kick and to 
result in a significant damping of the final BH motion. It is important
to recognize that in the scenario of Fryer \& Kusenko (2006) the
anisotropic neutrino emission carries momentum and that the same
amount of momentum in the opposite direction has to be {\em shared}
by the compact remnant and the SN ejecta.

A quantitative theoretical determination of the velocity distribution
of BHs in the fallback-SN scenario beyond the simple upper-limit estimate
of Eq.~(\ref{eq:escapevel}) will require long-time 3D hydrodynamic
explosion modeling including the fallback evolution for a large
set of progenitor stars. The gravitational tug-boat mechanism for NS
acceleration has been shown to lead to the
observational prediction that iron-group elements and explosively
produced intermediate-mass nuclei heavier than $^{28}$Si are ejected
by the SN with significant enhancement in the hemisphere of the 
stronger explosion, i.e., opposite to the direction of a large NS kick
(Wongwathanarat et al.\ 2013). In the case of BH formation in
anisotropic fallback SNe such an effect should even be enhanced. 
Strong hemispheric asymmetries of the heavy-element distribution in
SN remnants may thus provide observational hints for highly
aspherical explosions and therefore for large kicks of the compact
remnant, even if the latter and its space velocity might be 
difficult to measure.

A tight connection between BH kicks and SN explosion asymmetries 
implies that the BHs in fast-moving BH-LMXBs have to originate from
fallback SNe. BHs formed without accompanying SN explosions can be 
recoiled only by anisotropic neutrino emission, which might occur during
the neutrino cooling of a transiently stable NS. In this case the 
corresponding BH velocities would be limited by the value of 
Eq.~(\ref{eq:nukick-bh}), in conflict with the analysis by
Repetto et al.\ (2012). Accepting the inefficiency of neutrino-induced
BH kicks one can further conclude that stellar collapse to a BH without 
a SN explosion should lead to BHs with very low velocities, and
a bimodality of the BH velocity distribution seems possible. The 
relative population of the low-velocity and high-velocity components
should depend on the (still unknown) probability of BH formation without
associated SN relative to BH formation in fallback SNe.

\section*{Acknowledgments}

I would like to thank P.\ Crowther for a stimulating conversation
during the ESO Workshop ``Deaths of Stars and Lives of Galaxies'' in
Santiago de Chile, B.~M\"uller for a careful reading of the manuscript
and interesting comments, and M.\ Gilfanov and H.\ Ritter for sharing 
their knowledge of LMXBs with me.
Support by the Deutsche Forschungsgemeinschaft (DFG) through the 
Transregional Collaborative Research Center SFB/TR7 on ``Gravitational Wave
Astronomy'' and the Cluster of Excellence EXC153 ``Origin and Structure of 
the Universe'' (http://www.universe-cluster.de) is acknowledged.

\label{lastpage}
\end{document}